\documentclass[times,5p]{elsarticle}
\usepackage{lineno,hyperref}
\usepackage{physics, amssymb}
\usepackage{xcolor}
\usepackage{listings}
\usepackage{soul}
\biboptions{sort&compress}
\journal{Journal of Alloys and Compounds}

\begin{document}

\begin{frontmatter}

\title{
 Enhanced magnetoelastic stress in disordered iron-gallium alloy thin films revealed by direct measurement.
 }

\author[icma,myfootnote]{Adri\'an Begu\'e}
\author[icma]{Maria Grazia Proietti}
\author[icma]{Jos\'e Ignacio Arnaudas}
\author[icma,myfootnote1]{Miguel Ciria}
\address[icma]{Instituto de Nanociencia y Materiales de Arag\'on (INMA), CSIC-Universidad de Zaragoza, Spain.}

\fntext[myfootnote]{{Current address: Laboratoire Albert Fert, CNRS, Thales, Universit\'e  Paris-Saclay, 91767 Palaiseau, France}}
\fntext[myfootnote1]{Corresponding author: miguel.ciria@csic.es}

\begin{abstract}
The large magnetostriction in FeGa alloys is relevant for manifold applications, but for thin films, it can play a prominent role in controlling the strength of the magnetic anisotropy. Bulk samples show values depending on the extensive preparation procedure compendium, which is limited in its temperature range for high-quality thin-film synthesis. Here, we present a study of the magnetoelastic coupling coefficients  $B_1$  and $B_2$ in epitaxial FeGa thin films below 50 nm deposited on the MgO(001) surface at 150 $^\circ$C  by the cantilever method.
Series of films with 22, 28, and 33 at. \% Ga do not show thickness-dependent variations for $B_1$  and $B_2$, but $-B_1$ for the 22 at. \% Ga composition is 10 MPa, roughly 2 times the bulk value and smaller than the bulk-like value of $-B_1$=12.1 MPa obtained for a film with 17 at. \% Ga.  This enhancement is correlated with the A2 crystal structure for the film rather than the coexistence with D0$_3$ or other ordered nanometric precipitates proposed for bulk samples.
Synchrotron diffraction excludes the formation of long-range L6$_0$, or D0$_3$ precipitates in samples with (001)A2 peaks at concentrations around 25 at. \% Ga, which implies partial chemical disorder.  The analysis of extended x-ray absorption fine structure measurements points to a D0$_3$ local order with a residual number of Ga-Ga pairs. 
Considering that the substrate quenches the movable strain in the A2 phase described in dual-phase structures, our results point to the important role of the electronic structure of the iron atoms modified by the presence of Ga in the alloy. This effect enlarges $B_1$ in films with the A2 phase, stabilized using epitaxial growth.
\end{abstract}

\end{frontmatter}

\section{Introduction}
The Fe$_{100-x}$Ga$_x$ (x$<$35) alloy family displays compelling properties to be used in elements looking for energy-efficient applications. The compound with x $\approx$ 18 was first known for large values for the tetragonal magnetostrictive mode $\lambda_{100}$ \cite{Clark2001,Clark2003}, a property that coincides with small values for the lowest order magnetic anisotropy constant $K_1$ \cite{Rafique2004a}. 
 Its use in magnetoelectric heterostructured materials for voltage control of the magnetization based on strain-induced coupling between ferroelectric and magnetostrictive components has demonstrated large converse coupling, with coefficient values well above 10$^{-5}$ ms$^{-1}$ \cite{Begue2021,Meisenheimer2021}, a feature fundamental to reduce the energy consumption in processing information \cite{Hu2019} and neuromorphic computing \cite{Yang2020}.
On the other hand,  the ordered D0$_3$ phase (x $\sim$ 25) presents a stark giant anomalous Nernst effect at room temperature \cite{Sakai2020} suitable for recollecting energy using thermal variations due to the probable presence of a topological nodal-web structure \cite{Feng2024}.

\begin{figure*}
	\centering
	\includegraphics[width=1\linewidth]{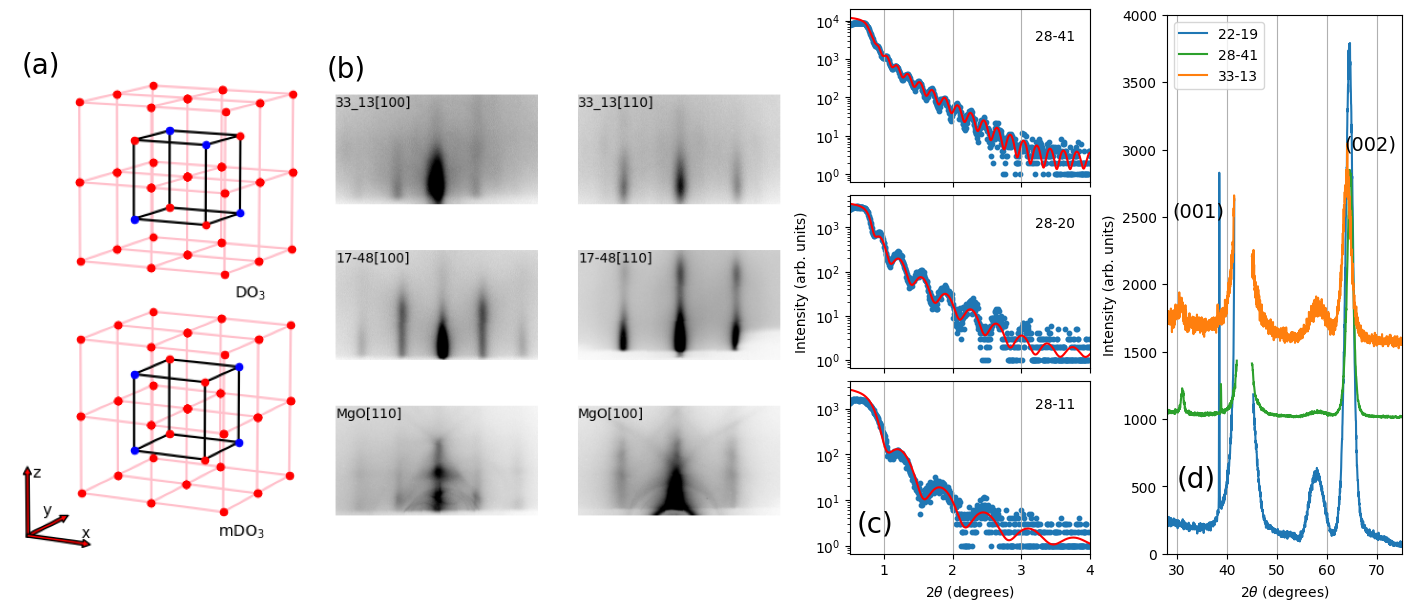}
	\caption{(a) Sketch of the mD0$_3$ and D0$_3$ structures,(b) RHEED images of the FeGa alloy and the MgO substrate before the film evaporation. (c) XRR data (dots) and fit (red line) performed to obtain the film thickness (d) Bragg-Brentano scans for films with and without the presence of the superlattice peak at 2$\theta \approx$ 30.6$^\circ$.}
	\label{g7}
\end{figure*}

The enhancement of  $\lambda_{100}$ in FeGa alloys with respect to the pure iron value has been explained using several frameworks. The idea of local order occurring by the generation of nanometric particles of a second phase \cite{Boisse2011, PALACHEVA2017229,Yangkun2018,ZHANG2022117594,YAN2024119583,He2016} or the electronic structure modification on the Fe atoms \cite{Wang2010,Wang2013, PhysRevB.103.094414, PhysRevB.99.054415,Adelani2021a,PhysRevB.109.014417} makes the FeGa alloys an archetype dual-phase systems. 
On one side, the presence of inclusions with a tetragonal symmetry, D$0_{22}$ and L6$_0$ (or mD0$_3$, see figure \ref{g7}a) in the cubic A2 matrix activates a strain in the matrix that enlarges the deformation in the single domain configuration \cite{Yangkun2018}. The role of the tetragonal inclusions as the seed of the enhanced magnetostriction with the rotation of their anisotropy axis by the magnetic field \cite{PhysRevLett.106.105703} or induced strain around the particles in the A2 matrix, which rotates and follows the magnetic field \cite{He2016,Yangkun2018}, are also under discussion. However, although the more significant increment of $\lambda_{100}$ with x $<$ 20 appears for samples obtained with a quenching from a high-temperature mix, $\lambda_{100}$ also takes values higher than that for iron for slow cooling metallurgic processes \cite{Xing2008}.
In any case, high photon flux experiments suggest that the volume fraction of mD$0_3$ is below 0.2\% due to the lack of pristine observation of the distinctive reflections \cite{PhysRevMaterials.8.073604}. Evaluating  $B_1$ by the indirect method of measuring the magnetic anisotropy suggests an enhancement associated with the A2 phase stabilized by epitaxial growing for x $>$ 20 \cite{Meisenheimer2021}.
 On the other side, extensive \textit{ab initio} studies  \cite{Wang2010,Wang2013, PhysRevB.103.094414, PhysRevB.99.054415,Adelani2021a,PhysRevB.109.014417} look into the modification of the Fe electronic state and its influence in the magnetoelastic (ME) coupling coefficient $B_1$ and stiffness elastic coefficients $c_{11}$, $c_{12}$,  $(3/2)\lambda_{100}=-B_1/(c_{11}-c_{12})$  as a function of the Ga content, and the presence of others 3d and 4f ions in the A2 phase, as well as the calculation of those coefficients for the B2 and D$0_3$ crystal structures. 

The complexity in finding a satisfactory explanation for the enhancement of $\lambda_{100}$ in Fe alloys reveals a rich behavior in samples prepared under different procedures that can not be replicated for thin films. Interestingly, in bulk samples, the density of those nano inclusions involves annealing at high temperatures (1000 $^{\circ}$C), quenching and aging at intermediate temperatures, and a second quenching \cite{NAsia}. These procedures are not accessible for thin film preparation aiming for flat topography and sharp interfaces since inter-diffusion with the substrate or buffer layers or the formation of rough layers with dome-like topography is activated by substrate temperature \cite{Begue2020}. In any case, in the nanometric range, FeGa crystalline films with low magnetic anisotropy show an ME stress large enough to activate uniaxial magnetic anisotropies \cite{Begue2021, Meisenheimer2021} by piezostrains and/or ferroelastic strain, with large converse magnetoelectric coupling coefficients. The magneto-crystalline barrier for moving M in the (001) plane between easy and hard direction is ($K_1$/4), which is below 10 kJm$^{-3}$ for x$>$18 \cite{Rafique2004a} and 1 kJm$^{-3}$ for x$>$25 \cite{10.1063/1.1856731}. It can be overcome if uniaxial strain as low as 0.001 is applied, considering that $-B_1$ can take values around 10 MPa, which is a value not unusual for thin films \cite{Meisenheimer2021,Barturen2019a} and about 4 times the Fe bulk value \cite{DUTREMOLETDELACHEISSERIE1983837}.

The effect of the epitaxial strain induced on the film introduces an environment different from that proposed for bulk samples. The induced strain by nano inclusions, if they nucleate,  may be concealed in very thin films: the internal strain is generated by the misfit with the substrate, and inhomogeneities are tied either to the relaxation of the strain through misfit dislocations of elastic deformation. The effect of the substrate can be released as the thickness of the film increases, as is shown by the increment of the temperature at which the lattice parameter is frozen by the substrate on Eu films \cite{Soriano2004}.
Therefore, the substrate restricts the rotation of distortions in the crystalline A2 matrix or the embedded nanoparticles for very thin films. The origin of the observed stress, which occurs when a magnetic field is applied and moves the magnetization away for the easy direction, highlights the significant role of an electron-like mechanism based on spin-coupling, primarily due to the introduction of gallium. That scenario is also different from the one observed in polycrystalline thin films: the effective ME coefficient $|B_{eff}|$ yields increasing \cite{JavedJAP} or decreasing \cite{JIMENEZ2020166361, Bartolome2020, PhysRevApplied.12.024020} values with increasing film thickness, that can be associated to modification of the crystal orientation of the grains or N\'eel surface contributions. Also, the strain-stress states of the grains in polycrystalline films induced by the substrate can be relevant in the observed value of $B_{eff}$ \cite{PhysRevB.104.064403}. 

Before dealing with the enhancement of $\lambda_{100}$ in FeGa alloys, it is worth mentioning the complexity of the magnetostriction phenomena of the single Fe element.  It is known that  $B_1$  shows a very unusual temperature dependence without decreasing $B_1$ as T rises from 0 K to room temperature. That departure from Akulov's law \cite{DUTREMOLETDELACHEISSERIE1983837}, which introduces the thermal disorder in the orientation of the magnetic moment, forced to consider shifts of the orbital levels as a function of the temperature to explain that behavior \cite{PhysRevB.99.054415, DOMINGUEZ1998121}. Also, a strain-sensitivity of the ME coupling coefficients $B_1$ and $B_2$ is demonstrated by the observation of decrements of the $|B_1|$ and $B_2$  bulk values for Fe films with thickness even above 100 nm grown on MgO(001) \cite{Koch1996, WEDLER2000896}.

Motivated by the strain-induced results observed in FeGa thin films,  the phenomenology associated with the local order chemical as well and the lack of measurements of the fundamental magnetoelastic coefficients for films below 50 nm, we report direct measurements of  $B_1$  and $B_2$  in epitaxial thin films grown on MgO(001) substrates, complemented with extended x-ray absorption fine structure measurements for the same set of samples. Therefore, measuring the \textit{B's} is of interest since the softening of the $c_{11}-c_{12}$ coefficient is the one reason for the observation of the second peak in the  $\lambda_{100}$ vs \textit{x} curve and also contribute partially to enhance $\lambda_{100}$ around \textit{x} = 19 \cite{Clark2003}.

\section{Materials and methods}

\subsection{Thin film preparation}

The procedure to obtain  Fe$_{100-x}$Ga$_x$ crystalline films onto MgO(001) crystals has been described elsewhere\cite{Ciria2018}. The film composition was set by co-evaporation of Ga and Fe using effusion cells and ebeam gun.
The films were grown at a substrate temperature of 150 $^\circ$C  and have thicknesses below 60 nm. The FeGa[100] in-plane direction is aligned with the MgO [110] axis.
Two rectangular MgO(001) plates with edges along [110] and [100] were used to obtain films suitable to measure $B_1$ and $B_2$, respectively, on films grown in the same batch. The structure of the layers is monitored \textit{in situ} using Reflection High Electron Energy Diffraction (RHEED), with electron beam energy equal to 15 keV, see Fig.\ref{g7}b. Continuous substrate rotation was enabled to improve the homogeneity of every layer.

\begin{figure*}
	\centering
	\includegraphics[width=1.0\linewidth]{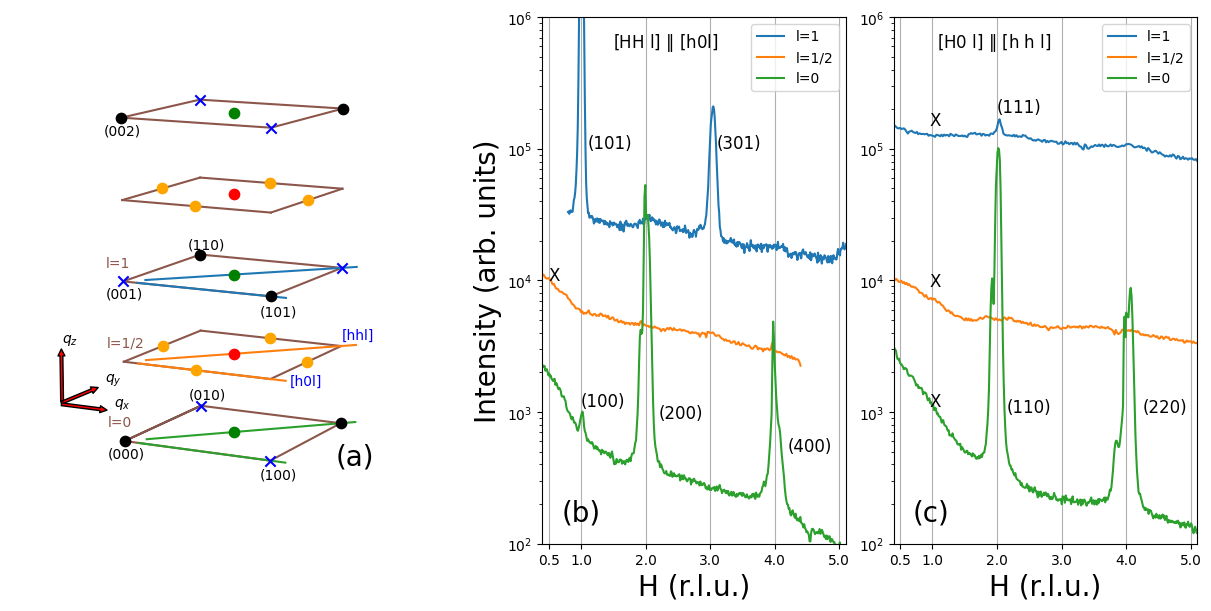}
	\caption{(a) Sketch of the reciprocal space for the A2 (black points) B2 (black points and crosses), D0$_3$ (black points, crosses, and red points) and mD0$_3$ (black points, crosses, and green points) with variants with Ga-pairs along the z Cartesian axes (Figure 1a).  For in-plane Ga-pairs, yellow points appear instead of green ones.  The reference axes correspond to the FeGa structure.  Reciprocal space linear scans along (b)  [HHl]$\|$[h0l] with l = 0, 1/2 and 1; and (c) along [H0l] $\|$ [hhl] with l = 0, 1/2 and 1 (b). The parameter H  is the Miller index of the MgO cubic lattice and the reciprocal lattice   has units of 2$\pi$/$a_{MgO}$. The labels  close to the peaks  indicate Miller indexes of observed FeGa reflections. Crosses are located at positions where extra reflections identifying the D0$_3$ or mD0$_3$ structures should have been observed.}
	\label{Gren}
\end{figure*} 

Three sets of films with x $\approx$ 22, 28, and 33 and film thickness range between 10 nm and 50 nm were studied, as well as several films with thickness around 50 nm and varying Ga concentration.  Table \ref{list_films} shows the relation of films studied in this work and some relevant parameters such as thickness and estimated composition obtained by energy-dispersive X-ray spectroscopy (EDX) and  Wavelength Dispersive Spectroscopy (WDS) in an electron probe micro-analyzer (EPMA) instrument.

\subsection{Characterization techniques}

XRD diffraction was performed in the Bragg-Brentano configuration with a RIGAKU Ru2500 diffractometer (copper anode 40 kV/80 mA, with $\lambda$= 1.54 \AA)   and synchrotron radiation  (BM25 Spain Beamline at ESRF, with $\lambda$= 0.62 \AA). X-ray reflectometry was used to obtain the film thickness by fitting the intensity oscillations using a Bruker D8 diffractometer, Fig.\ref{g7}c. 
X-ray Absorption Fine Structure (XAFS) spectroscopy is a short-range-order, chemically selective technique complementary to diffraction. It can provide information on the atomic local environment of the different atoms comprising the samples (see ref  \cite{Lahiri2024} for an overview).
The XAFS spectra $\mu(E)$  were recorded with the beam polarization nearly parallel  ($\epsilon_{\|}$,  incidence angle  $\cong$ 5$^\circ$) and almost perpendicular ($\epsilon_{\bot}$,  incidence angle  $\cong$  85$^\circ$) to the sample surface. We measured the extended region above the absorption edge of the XAFS spectrum (EXAFS) and the Near-Edge region (XANES)  at the Fe and Ga K-edges at beamline B18 of the Diamond UK’s National Light Source. 
The raw polarized XANES spectra with parallel and perpendicular polarization were subtracted from each other to obtain the Linear Dichroism XANES signal (LDXANES) \cite{deGroot2006}. 
 
Magnetic hysteresis loops were obtained by using a magnetooptical Kerr effect (MOKE) magnetometer (nanoMoke3) with the field \textit{H} in the film plane, see Figure \ref{fig:VSM}. Also, from the hysteresis loops the energies requested to saturate the film are estimated by evaluating the area \textit{A}($\phi$) of the anhysteretic curve  and the bulk magnetization  value $M_s$. Thus, 
the lowest order magnetic anisotropy constant $K_1$ is calculated as A([110])-A([100])=$K_1/4$. 
All the films were studied using the cantilever method with a magnetic field rotation of about  $\mu_0 H=$100 mT. The rotation of the field is required to extract the magnetization \textbf{M} from the easy direction since if $\mu_0 H$ is applied along the easy axis, the inversion of M takes place suddenly at the coercive field but without strain change, since a 180$^\circ$ domain wall movement can take place to switch M and $\lambda \propto M^2$. See the appendix for details relating the magnetoelastic stresses with the cantilever capacitive detection method.

\begin{table}
\centering
	\	\begin{tabular}{ccccc}
	\textit{Sample} &\textit{thickness} (nm) &  \textit{x} ($\%$ Ga)& $-B_1$ (MPa) & $B_2$(MPa)\\
		\hline 	       
		\hline 
		Fe     & 73   & 0   & 2.1 & 5.5  \\
		17-48  & 48   & 17.2& 12.1  & 6  \\
		22-10  &10.6  &22   & 9.5 & 1.5  \\
		22-19  &19.0  &22  & 10.3 & 1.6  \\
		22-33  &32.6  &22   & 10.3 & 1.3  \\
		28-11  &10.8  &28   & 6.1 & -5.3  \\
		28-20  &20.0  &28   & 5.9 & -4.8  \\
		28-31  &31.8  &28   & 7.4 & -4.8 \\
		28-41  &41.1  &28   & 6.5 & -5.3 \\
		33-13  &13.0  &33   & 5.9 & -3.9 \\
		33-25  &25.0  &33   & 6.2 & -3.9  \\
		33-36  &35.7  &33   & 7.4 & -3.5  \\
		26-52  &52    &26   & - & -4.4  \\
	    \hline
	\end{tabular}
	\caption {Set of films measured in this work. The code identifies samples by composition x and thickness. The values of  $B_1$ and $B_2$ are also indicated.}
	\label{list_films}
\end{table}

\section{Results and discussion}

\subsection{X-ray diffraction characterization}

XRD scans in the Brag-Brentano configuration show the FeGa (002) and the Mo (002) reflections around 2$\theta \approx$ 64$^\circ$ and 58$^\circ$, respectively, and above x = 22 the (001) one at 2$\theta \approx$ 30.6$^\circ$, see figure \ref{g7}d. The latter weak reflection indicates ordering between the Fe and Ga species, and it is common to B2,  D0$_3$, and mD0$_3$ structures. Those weak peaks are labeled as superlattice (SL) reflections to distinguish them from the strong reflections originating from the A2 structure. The corresponding out-of-plane lattice parameters,
 obtained from the (002) reflection, are between 2.88 \AA\ and 2.91 \AA, which are values smaller than those observed for bulk samples with the same composition \cite{Kawamiya1972}. Thus, see Figure 1d,  samples 22-19 ($2\theta$ = 64.58$^o$), 28-41 ($2\theta$ = 64.68$^o$) and 33-47 ($2\theta$ = 63.94 $^o$) yield 2.88 \AA, 2.88 \AA\ and 2.91 \AA\ respectively.  The bulk value for x=23 is about 2.91 \AA. The (001) peak is observed at $2\theta$ = 31.116$^o$ for the 28-41 film and $2\theta$ = 30.58$^o$  for the 33-47 one.  The films studied here do not show the presence of phases with the fcc-like arrangement of the atoms observed by the presence of the (002) reflections around 2$\theta \approx$ 49$^\circ$ (a$\approx$3.69 \AA) \cite{Ciria2018, Kawamiya1972}. 

X-ray diffraction employing synchrotron radiation was used to look for SL reflections characteristic of D0$_3$ and mD0$_3$ structures, see Figure \ref{Gren},  in the 26-52 film, with the (001) reflection observed in the  Brag-Brentano configuration.
Here, we use the notation h, k, and l for the Miller indices for the  A2 reciprocal space lattice, space group $Im\overline{3}m$, Z = 2 and unit vector 2$\pi$/$a$, with $a$ the lattice parameter to be obtained but around 2.9 \AA. The B2 structure, space group $Pm\overline{3}m$ implies an order between Fe and Ga in the cell and many Ga-Ga pairs.  For the D0$_3$ and mD0$_3$, the cell unit is doubled in real space, see Fig.\ref{g7}a,
and ordering appears with fractional values of h, k, and l \cite{Du2010}. Using the D0$_3$ cell avoids the fractional notation and is straightforwardly done by multiplying by two the Miller indices.
Figure \ref{Gren}.a shows a sketch of the reciprocal space with the reflections due to the A2 phase (h+k+l = even, black points) and SL reflections. For B2 crosses are added at h+k+l = odd; for the  D$0_3$ phase distinguished reflections are added at  [1/2](h k l), the red points, while for mD0$_3$ those specific SL peaks correspond to the green points at (h+0.5 k+0.5 l)  for pairs along the [001],  and (h+0.5 k l+0.5) or (h k+0.5 l+0.5) for in-plain  [010] and [100]  Ga pairs, see orange points.  The lines indicate scans performed with synchrotron radiation to ascertain the presence of the D0$_3$ or mD0$_3$ phases and presented in Fig\ref{Gren} a and b. Those scans use the MgO crystal to set the reference frame (with H, K, and L Miller indices and 2$\pi$/$a_{MgO}$ modulus of the unit cubic vector). The epitaxy between the MgO [110] parallel to the FeGa [100] of the \textit{bcc} structure is observed at the scans with L $\approx 0$, where peak overlapping is manifested, see  Fig.\ref{Gren} b and c. Therefore, for scans along  [HH.], h $\approx$ H, while for [H0.]  h=k $\approx$ 2H. The more intense peaks correspond to the Bragg condition for the \textit{bcc} lattice (h+k+l = even).
Thus, [HH.] is parallel to [h0.], and the value of L for the FeGa reflection (l=1,2, etc.) is determined by maximizing the intensity for the FeGa (002) reflection.
 For l=1, blue lines, the A2 (101), (301), and (501) reflections, and the SL peak (111) are visible.

The scans [hh 1] (blue line) and [h0 1/2] (orange line) did not show  (1/2 1/2 1)  and  (1/2 0 1/2)  contributions from mD0$_3$ with [001] and in-plane Ga-pairs, respectively. On the other hand, the scan [hh 1/2] (orange line) discards the  D$0_3$  phase by the absence of the  (1/2 1/2 1/2)  peak.  The scans at grazing incidence (l $\approx$ 0, green lines) only show the  (100) peaks but not  (1/2 1/2 0)  corresponding to the mD$0_3$ with [001]Ga-pairs.
The lack of SL peaks in epitaxial alloy films has also been observed in Heusler compounds. It suggests the presence of chemical disorder in the structure \cite{Heusler}.  Therefore,  (001) planes of Fe atoms sandwiched between a single atomic layer of a disordered mixture of Fe and Ga atoms generate a B2-like structure that excludes the formation of long-range D$0_3$ or mD$0_3$ structures (see Fig. \ref{g7}a). 
Interestingly, the growth at 150 $^{\circ}$C shows the well-known trend of Ga to avoid other Ga as the first neighbor. However, at that temperature, the formation of long-range order in the FeGa layer is inhibited.  
These scans allow determining the  out-of-plane and in-plane  lattice parameters resulting in a value of 2.88 \AA\ and 2.93 \AA\ respectively, and demonstrating a tetragonal deformation of the cubic cell enlarging the in-plane distances due to the epitaxy with the in-plane distance of the MgO substrate ($a_{MgO}$ = 4.212 \AA, $\sqrt{2}a_{FeGa}\approx$ 4.1 \AA)

 \subsection{EXAFS Results}
 
 The experimental spectra were fitted by theoretical EXAFS signals calculated by {\it ab initio} theoretical phase and amplitudes generated for a cluster of 78 atoms with a radius of 6.1 {\AA}$^{-1}$, by the TKAtoms code \cite{Ravel2001}. The central absorber was a Ga or Fe atom with only one kind of Ga (Fe) scatterer. The alloying FeGa effect was obtained by combining the pure Ga-Fe(Ga) and Fe-Fe(Ga)systems at the Ga and Fe K-edge, respectively. The 8.01 Feff code generated the theoretical phase-shifts and amplitudes \cite{Ankudinov1998} taking into account beam polarization. The  DEMETER package was  used for extraction and best-fit procedure of the EXAFS spectra \cite{Ravel2005}. 

\begin{figure}
    \centering
	\includegraphics[width=1\linewidth]{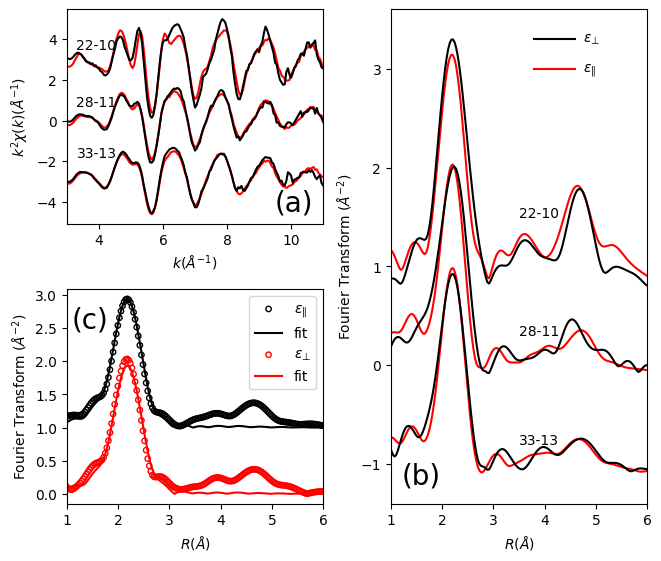}
	\caption{a) Raw background subtracted EXAFS spectra and b) correspondent Fourier Transform amplitudes at the Ga K-edge, of some of the samples studied with close to in-plane (red line) and out-of-plane
(black line) polarization of the incoming beam. An example of best-fit results is given in panel c), showing FT amplitudes of the EXAFS signal (circles) for the two X-ray beam
polarization directions, for one of the samples (33-13) together with the
corresponding best-fit curves (solid lines). Fits were performed in q-space.}
	\label{fig:EX1}
\end{figure}

The raw EXAFS spectra were Fourier Transformed (FT) in the range 2 $<$ k $<$  11 \AA$^{-1}$. 
Figure \ref{fig:EX1}a shows raw background subtracted EXAFS spectra at the Ga K-edge, and Fig. \ref{fig:EX1}b, the correspondent FT signals, for the two beam polarization directions.
The first peak of the FT spectra is due to the first and second coordination shells. The contributions of next neighbor (NN) and next-nearest-neighbor (NNN) atoms are merged into a single FT peak due to the proximity of the interatomic distances values (2.48\ \AA\ and 2.87\ \AA\ respectively in pure bcc Fe)
 and the limited transformation range. 
The best-fit analysis  was performed simultaneously  for the two polarizations in q-space,  by Fourier Filtering of the first peak contribution in the range 1 $<$ R $<$  3 \AA. The best-fit parameters were: the origin of the photoelectron energy $E_{0}$, the Ga population in NN and NNN environments, $x_I$ and $x_{II}$, the interatomic distances of first, $d_{I}$, and in-plane second $d_{II\|}$ coordination shell, and the Debye-Waller factors. The $S_0^2$ amplitude reduction factor was set to 0.7 for all the samples. 

The $E_{0}$ value was refined for one
of the samples and found equal to 1.5 eV. It was kept
fixed to this value for the rest of the samples. We
performed the fits by fixing the coordination numbers to
the crystallographic values for bcc Fe (8 for the NN atoms and 6
for the NNN neighbors) and letting the Ga population vary
together with the Debye-Waller factors and interatomic
distances. The variation of the interatomic distances was
limited in a range physically reasonable according to our
previous results on the same system.

 More details about the EXAFS analysis are given in a previous Galfenol paper dealing with the same kind of samples \cite{Ciria2018}.
 Figure \ref{fig:EX1}c displays an example of the best-fit curves in R-space for the two polarizations for one of the samples presented here. The overall results are reported in Table \ref{tab}. 

\begin{table*}
	\begin{tabular}{c|ccccccc}
		\textit{Sample} &  $x_{I}$ & $x_{II}$  & $\sigma^2_{I}$ (\AA${^2}x10^{-3}$)&$\sigma^2_{II}$ (\AA${^2}x10^{-3}$)&   $d_I$ (\AA) & $d_{II\|}$ (\AA)& $d_{II\bot}$ (\AA) \\
				
		& \textit{K-Ga\;  K-Fe} & \textit{K-Ga\;  K-Fe }   & \textit{K-Ga\;  K-Fe}  &   \textit{K-Ga\;  K-Fe} &   \textit{(Ga-Fe) \;(Fe-Ga)}   & \textit{(Ga-Fe)\; (Fe-Ga(Fe))} &  (\textit{diff}) \\
		\hline 
		22-10        &0.0(1)\; 0.32(7)  &0.0\;  0.0  & 7(1)\; 4.5(6) & 16(2)\; 16(3)&2.52(1)\;  2.53(2)  & 2.87(1)\; 2.89(2)& 2.89\\
		
		22-33          &0.0(1)\; 0.36(6)  &0.0\; 0.0    & 7(1)\; 5(1) & 19(2)\; 15(3) & 2.52(1)\; 2.53(2) & 2.87(1)\; 2.87(2) & 2.89\\
		
		28-11          &0.0(1)\; 0.47(4)  &0.0\; 0.0    & 10(1)\; 6(1) & 27(1)\; 30(6)&2.52(1)\; 2.53(2)  & 2.87(1)\; 2.87(2) &2.89\\

		28-41        & 0.0(2)\;  0.56(4) &0.0\; 0.0   &  8(1)\; 7(1) & 28(1)\; 38(8)&2.51(1)\; 2.53(1) & 2.90(1)\; 2.87(3) &2.89\\
		
		33-13        &0.0(2)\;  0.56(4)   &0.0\; 0.0   & 9(1)\; 6(1) & 30(9)\; 44(15)&2.53(1)\; 2.53(1)  & 2.89(1)\; 2.88(4) &2.91\\
		
		33-25         &0.1(2)\;  0.51(4)  &0.0\; 0.0    & 9(1)\; 6(1) & 27(9)\; 32(7)&2.50(1)\; 2.53(3) & 2.87(2)\; 2.89(3) &2.91\\
		
		33-36      &0.0(2)\;  0.56(4)  &- - &  10(1)\; 7(1) & 40(12) \;39(9)&2.52(1)\; 2.53(1)  & 2.87(1)\; 2.87(3) &2.91\\
		
		\hline %
	\end{tabular}
		\caption {Best fit results for the studied samples at the Ga and Fe K-edges. The perpendicular and parallel polarization spectra were fitted simultaneously. We report the fit parameters as Ga
			population I and II shells ($x_{I}$ and $x_{II}$),  Debye-Waller
			factors,
		interatomic distances of I coordination shell, $d_I$, and II coordination shells, in-plane $d_{II\|}$, and out-of-plane, $d_{II\bot}$. The statistical
errors, given in parenthesis, were calculated from the diagonal values of the fit covariance matrix. The errors on $x_{II}$ are all about 0.1. The interatomic out-of-plane distance values for the II coordination
		shell were kept fixed to the values found by
		diffraction. }
\label{tab}
\end{table*}
All the samples show the same local environment around the Ga absorber at the Ga K-edge, independent of Ga concentration and sample thickness: no gallium atoms are found in the NN and NNN of the Ga absorber. That is expected for a D0$_3$-like ordering at a local scale and excludes the mD0$_3$ structure.
When we look at the Fe environment, by performing EXAFS at the Fe K-edge, we found a population of Ga scatterers, in the NN coordination shell close to 30\%  for the 22\% concentration, and close to
50\% for the 28\% and 33\% Ga samples. No Ga atoms are found in the NNN shell. This is in agreement with what is  observed at the Ga K-edge, i.e., the general and expected tendency of the Ga atoms to stay apart from each other at the local scale \cite{Ciria2018}.   
Short-range D0$_3$ ordering occurs, in which 50\% of the NN are Ga atoms and no Ga NNN are expected.  We have to note that the precision in determining composition by EXAFS is not very high, and we cannot exclude the presence of Ga as a scatterer below values of up to 10\% in atomic composition.
The interatomic distances of the first shell are determined for Ga-Fe (Ga threshold), Fe-Ga
(Fe threshold), and Fe-Fe (Fe threshold), and exhibit a bimodal distribution. The Fe-Fe
distance is equal to the bulk Fe value, while the Ga-Fe distance is around 2.52 \AA, as
previously observed in the literature \cite{Pascarelli2008}.
The determination of the interatomic distances of the second shell is more challenging
and less precise. Diffraction measurements provide the value of the lattice parameter in
the direction perpendicular to the surface, which is used in the fits to determine the in-plane
lattice parameter. A slight distortion is observed in the samples with x = 22 and 28, and a somewhat higher distortion in the x = 33 samples.
An important increase in the structural disorder is observed as the Ga content increases, both
in the Ga and Fe environments.
No structural changes are observed as the sample thickness increases.

 \subsection{XANES Results}
\begin{figure}
	\centering
	\includegraphics[width=1\linewidth]{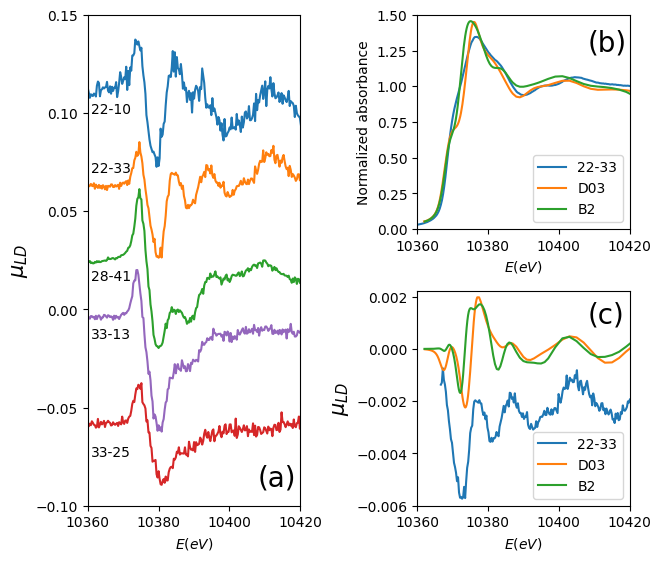}
	\caption{a) LD-XANES spectra at the Ga 
		K-edge, obtained by subtraction of in-plane and out-of-plane polarized XANES spectra for the studied samples.  b) XANES spectra, with in-plane polarization, for sample 22-33 (blue
curve) are compared with two \textit{ab initio} simulations, performed by the FDMNES code, for two clusters
with 9 \AA\ radius, having D0$_3$ (yellow curve) and B2 (green curve) symmetry. c) LD-XANES signal for sample 22-33 compared with the two simulations, D0$_3$ (yellow curve) and B2 (green curve).}
\label{fig:EX2}
\end{figure}

We have also studied the near-edge region of the XAFS spectra, which is known to be sensitive to small changes in the geometry of the absorber environment. The Ga-Ga pairs presence has been addressed
in previous papers by analyzing the Linear Dichroism signal in the XANES
spectra at the Ga K-edge \cite{Ranchal2018, Bartolome2020}.
 XANES is more sensitive  than EXAFS to local geometrical distortions  which gives rise to anisotropy. 
The XANES raw spectra recorded with in-plane and out-of-plane polarization were normalized to each other in the post-edge extended region and the LD Xanes signals were calculated as $\mu_{LD}=\mu_{\parallel}-\mu_{\perp}$. They are shown in Fig. \ref{fig:EX2}a for the samples studied with the exception of samples 28-11 and 33-36, the spectra of which did not have enough quality to perform this analysis.   
A dichroic signal of the same type 
is observed in all the samples, as expected due to the distortion of the cubic cell. The LD  signal is more intense for the samples with 21\% Ga, which do not show a SL  diffraction peak. On the other hand, according to EXAFS, they are characterized by a more ordered local environment than the richer Ga samples.
The observed  $\mu_{LD}$ agree with a previous paper reporting on the origin of magnetic anisotropy of Ga-rich FeGa films grown by sputtering with different thicknesses\cite{Ranchal2018}. In that study, a feature was observed that could be ascribed to the formation of Ga-Ga pairs only for thicknesses above 100 nm. 
Our samples are thinner than 50 nm and do not show such a feature.
A qualitative interpretation of the XANES spectra can be provided by performing \textit{ab initio} simulations and comparing the overall theoretical spectrum with the experiment. 
We used the FDMNES code \cite{Joly2009} 
for \textit{ab initio} calculation of the X-ray absorption threshold in the two polarization directions and the correspondent $\mu_{LD}$.  The simulations were performed for  an FeGa cluster with D0$_3$  (25 \% Ga) and B2 (50 \%
Ga) ordering. A 1\%\ distortion of the crystal lattice was applied while
keeping the cell volume constant. We verified that the calculation converges for a 9\ \AA\ cluster.
The results are shown in Figure \ref{fig:EX2}b, where we compare one of the experimental XANES spectra with the theoretical XANES for the two models, and in Figure \ref{fig:EX2}c, in which we compare experimental and theoretical $\mu_{LD}$. The theoretical XANES spectra qualitatively reproduce the experimental spectra, which appear to have a shape between the D0$_3$ and B2 structure calculations.

The XANES results are consistent with the EXAFS results, showing the tendency of Ga atoms to be apart from each other but without excluding the presence of a low number of Ga-Ga pairs. 
 If we compare the $\mu_{LD}$ of the whole sample set, we can observe a clear lowering of the $\mu_{LD}$ intensity when increasing the Ga content. This can be associated with the increasing local disorder observed by EXAFS. It could, in turn, favor the formation of pairs, which, even in small numbers, can modulate the magnetic behavior of the system.

\subsection{Magnetic anisotropy}

\begin{figure}
	\centering
	\includegraphics[width=1\linewidth]{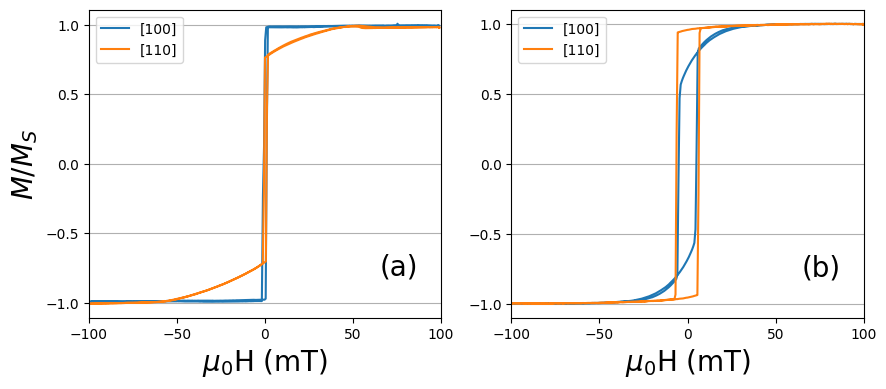}
	\caption{M-H loops with H along the [100] and [110] directions for (a) Fe film and (b) 22-19 film.}
	\label{fig:VSM}
\end{figure}

Figure \ref{fig:VSM} displays hysteresis loops obtained by MOKE, with the signal normalized to the saturation value,
and $\mu_0 H$ applied along the [100] and  [1{1}0] axis. It illustrates the behavior of the three compositions presented here comparing a reference Fe film and sample 22-19. 
Thus, for the pure iron films, the easy direction (EA) is the $\langle100\rangle$ axes, while for the  FeGa films, the EA shifts to the  $\langle110\rangle$ in-plane direction. $K_1 \approx$ -16 kJm$^{-3}$ for the  22-19 film, $M_s  \approx $ 1.45 T  and around 30 kJm$^{-3}$ for a reference Fe film,  $M_s \approx $ 2.08 T \cite{Restorff2012}.

\begin{figure*}
\centering
	\includegraphics[width=1\linewidth]{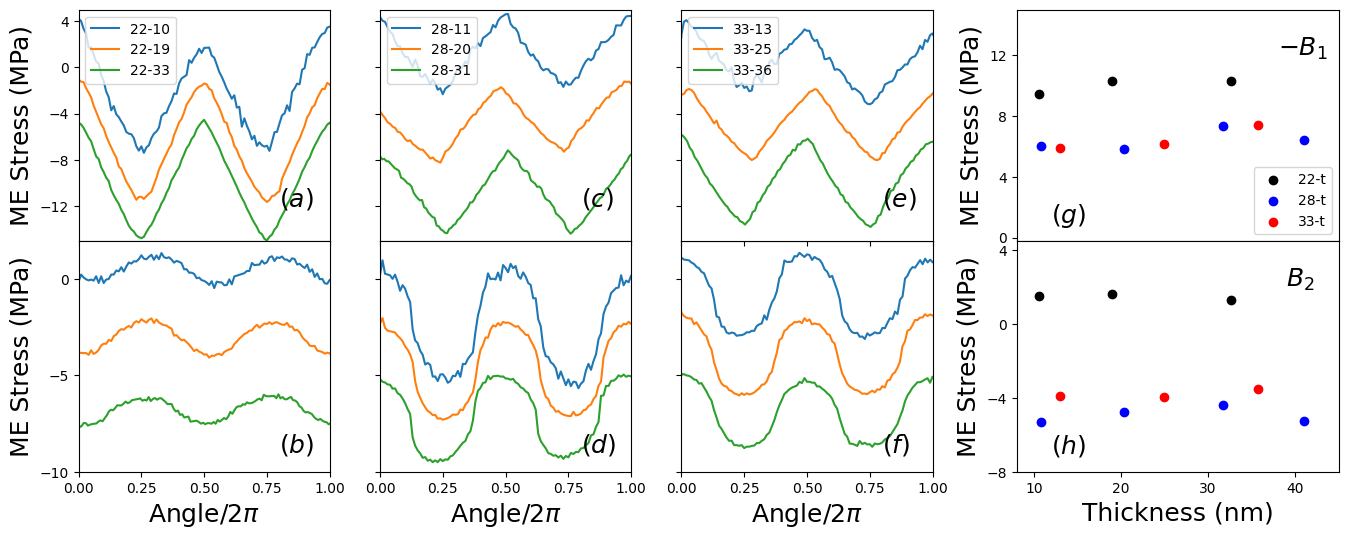}
	\caption{Magnetoelastic stress measurements vs angle (a)-(f)  for films with  x = 22 (a) and (b), x = 28 (c) and (d) x =33 (e) and (f). The angle = 0 position is set with $\mu_0H \|$ beam length.   Panels (a), (c), and (e)  show measurements performed in films with beams along the FeGa[100] direction to determine B$_1$. For
	panels (b),(d), and (f)  beams is parallel to FeGa[110] obtaining B$_2$. Stress curves are shifted along the y-axis for clarity purposes. Thickness dependencies for  the (g) $-B_1$ and (h) B$_2$ ME coefficients.}
	\label{fig:B1}
\end{figure*}
\subsection{Magnetoelastic stresses}
Figure \ref{fig:B1}a-f shows sets of ME stress curves for samples of the   three compositions presented in this work. For the thinner films, around 10 nm, they present jumps due to the lower signal-noise ratio, which is proportional to the film volume. The shape of these curves reflects the in-plane easy axis orientation [110], with a saw and parabolic features  when the hard and easy axis are along the cantilever beam. It is noted that the films are saturated at 100 mT as is shown in Fig. \ref{fig:VSM}.

The thickness dependence of $B_1$ and $B_2$ is displayed in Figure \ref{fig:B1} g and h. The variation of values is minute and within the experimental error, including composition fluctuation. The latter may be more relevant for  $B_2$  the x = 22 series since it changes sign around this composition. 
The largest value of $-B_1\sim$10 MPa in those sets of films is observed for x = 22 and decreases to $-B_1\sim$6 MPa for higher concentrations of Ga. $B_1$, in modulus, is larger for all compositions than that reported for Fe bulk ($-B_1$ = 3.44 MPa).
The highest value is obtained for a film with 17-48, 12 MPa, the composition where the first peak of  $\lambda_{100}(x)$ takes place. Interestingly,  $-B_1\sim$ 10 MPa obtained in the films with x$\approx$22 is about two times larger than the bulk value, about 5 MPa, calculated from the $(3/2)\lambda_{100}= $200 $\cdot$ 10$^{-6}$  and $c_{11}-c_{12} =$ 25.2 GPa  \cite{Restorff2012}.  Above that composition, $-B_1$ obtained here take similar values to the bulk ones, where $-B_1=$ 6.5 MPa for x=28.8 and 9 MPa for x =35.2 \cite{Restorff2012}.
EXAFS results, with spectra taken at the Ga K-edge support the absence of local ordering or a significant pairs between the reference Ga atoms and the positions at neighbor first of second layers, with a minute presence of Ga-Ga NNN pairs.  The presence of some Ga-Ga pairs is suggested qualitatively by the analysis of $\mu_{LD}$  by introducing the B2 phase in the environment of the Ga atom.  

Because of the strong link between crystal structure and magnetoelastic stress in FeGa alloys \cite{Xing2008, Du2010}, we note that the enhancement of  $B_1$  is observed with the absence of SL peaks in x-ray experiments, see Figure \ref{g7}.d where the (001) reflection does not appear for the 22-19 film indicating that only the A2 phase is formed in the film. For the x = 28 and 33 compositions, the (001) peak is displayed. Therefore, this experimental fact points to values for $B_1$ in films with the A2 crystal structure that multiply by 2 the bulk value. 
The direct observation of a large value of $-B_1$ in films with A2 structure agrees with the observation obtained in films grown on PMN-PT, where a significant increment of $-B_1 \approx$  18 MPa was obtained by the measurement of magnetic anisotropy constant \cite{Meisenheimer2021}. 
 \textit{Ab initio} calculations with the quasirandom-structure method to face disordered supercells \cite{PhysRevB.109.014417}, support the enhancement of $-B_1$(=8.7 MPa)  with respect to  Fe calculated value (=5.1 MPa) by the presence of random Ga atoms in an A2 structure with x=18.75. That increment for $-B_1$ is larger if Ga-Ga pairs are introduced in the calculation with the mD0$_3$ structure (=19.2 MPa).  
In bulk samples, the preparation procedure involves the coexistence of A2 and other phases, with a proportion that depends on the recipe used. The preparation method for flat thin films, growth at a substrate temperature of 150 $^{\circ}$C, and low flux of materials may retard the formation of a more ordered phase in favor of the disordered  A2 structure. 

$B_2$ is smaller, in modulus, than the corresponding bulk values \cite{Restorff2012} obtained for the film studied here, including the Fe film. Thus, bulk values are -12  MPa for x = 28 and -9.2 MPa for x = 33, while thin films are roughly forty percent of the bulk for those compositions.  On the other hand, for x = 22, $B_2$ is positive while the bulk value is negative, indicating that the switch of the sign of $B_2$ is shifted toward higher x values due to the persistence of the A2 structure in the thin films.

 Another observation regarding $B_1$ and $B_2$ 
 in FeGa films is the lack of significant effects associated with film thickness,  contrasting with the behavior reported for Fe films.
 This result  may  agree with the conclusion obtained from the fully relativistic disordered local moment theory applied to
fully disordered  FeGa alloys \cite{PhysRevB.99.054415}. In this study, increasing the Ga content reduces the  volume sensitivity of $B_1$ calculated for pure iron, becoming it negligible for x = 20, although the same calculation does not show enhancement of $B_1$ for disordered A2 FeGa alloys.

Finally, we hypothesize about the role of the intrinsic deformation of the A2 phase. 
The lattice parameters data indicate that the films undergo a tetragonal distortion with the enlarged azimuthal plane. Extrinsic mechanisms require the rotation of the strained A2 phase with local distortions by nanoparticles with Ga-Ga pairs along the magnetic field \cite{Yangkun2018}, so only mD0$_3$ particles within-plane Ga-Ga pairs would contribute with $\mu_0H$ rotating in the film plane.  However, the distortion of these particles and the A2 matrix is affected and controlled by the substrate, which promotes lattice parameters either with homogeneous deformation or, above a certain critical thickness, a relaxation by generating misfit dislocations.  Therefore, local internal stresses can be large  because they are proportional to the difference between actual and unconstrained lattice parameters, but tied to the substrate. Thus, the substrate would freeze the displacement of the induced deformation in the A2 structure by uniaxial particles, and only the distortion generated by the spin-orbit effect remains. It is noticeable that even in films with SL peaks, synchrotron X-ray diffraction, and EXAFS discard the presence of uniaxial particles in significant proportions.

\section{Summary}

In this article, the tetragonal and rhombohedral magnetoelastic coupling coefficients are obtained in epitaxial thin films as a function of composition and thickness using a cantilever method. We show the enhancement of $-B_1 \approx$ 10 MPa in thin films with A2 structure, at the composition with x = 22,  which presents the D0$_3$ precipitates in bulk samples, doubling that bulk value. The direct observation of a significant value of $-B_1$ in films with A2 structure jibes with $B_1$ obtained through the measurement of the modification magnetic anisotropy constant with strain in films grown on ferroelectric PMN-PT substrates \cite{Meisenheimer2021}. 
The enhancement of $-B_1$ due to the presence of random Ga atoms in an A2 structure (x=18.75) is also supported by \textit{ab initio} calculations with the quasirandom-structure method to face disordered supercells \cite{PhysRevB.109.014417} being that increment larger if Ga-Ga pairs are introduced in the calculation (mD0$_3$ structure).

The dependence of $B_1$ and $B_2$  with the film thickness does not show variations associated with that parameter.  In addition, 
$-B_1$ has a maximum for x=17 and decreases for any compositions with higher content of Ga.
Regarding $B_2$, smaller values than bulk values are obtained for the films studied here, including the Fe film. The switch of the sign of $B_2$ in the films is shifted toward higher x values due to the persistence of the A2 structure.

Thus, we suggest that the strain induced by the substrate in combination with the \textit{low-temperature} growth conditions hinder the formation of ordered species of the FeGa alloy (D0$_3$,  mD0$_3$, B2, etc.) observed or proposed in bulk samples, with the interesting result of the enlarging $B_1$ by a factor 2 compared to the bulk value for x = 22. Therefore, taking 
K$_1$/ 4 $\approx$ 4 kJm$^{-3}$ for the 22-19 film, the condition  $B_1 \epsilon > K_1/ 4$ to control energy minima with strain switching may require $| \epsilon |$ as low as 0.4 $\cdot$ 10$^{-3}$  to revert the sign of the uniaxial anisotropy and reduce the strength of external stimuli related to $\epsilon$ applied to magnetic devices placed at the central gap of electrode pairs on ferroelectric substrates \cite{Cheneaay5141}.
Therefore, we showed that the epitaxy of FeGa with large ME stress and low magnetoelastic anisotropy can be obtained in thin films with low roughness without needing high-temperature treatments.  

FeGa and other heterostructured materials open the door to improve magnetoelastic performance by several mechanisms that can be complementary to each other; one is the modification of electronic structure and the activation of soft phases by inclusions and nanoprecipitates. From a practical point of view, these short-range interactions boost the magnetoelastic stress and diminish the magnetic anisotropy with applications at low fields and malleable magnetic anisotropy by applying external strain. In conclusion, we infer that the strength of the electronic effects of \textit{bcc} iron alloys is worth exploring and suggest the relevance of the modifications of the  iron atom electronic structure  by alloying with gallium. 

\section{CRediT authorship contribution statement}

\textbf{Adrian Begu\'e}: data curation, formal analysis, investigation,  validation, writing - review \& editing.
\textbf{Maria Grazia Proietti}:  data curation, formal analysis, visualization investigation,  validation, writing - review \& editing,  funding acquisition.
\textbf{Jos\'e Ignacio Arnaudas}: formal analysis, writing - review \& editing,  funding acquisition.
\textbf{Miguel Ciria}: data curation, validation, investigation, formal analysis, visualization, funding acquisition,
writing – original draft.

\section{Acknowledgments}

We thanks grant TED2021-131064B-I00 funded by \\
MCIN/AEI/10.13039/501100011033 and by European Union NextGenerationEU/PRTR, grant
PID2021-124734OB-C21 fun-ded by MCIN /AEI /10.13039/501100011033  and by “ERDF A way of making Europe” and grant CEX2023-001286-S funded by MICIU/AEI /10.13039/501100011033 and Gobierno de Aragón (grant E12-23R). A. B. would like to acknowledge the funding received from the Ministry of Universities and the European Union-Next Generation for the Margarita Salas fellowship. We are grateful to the staff of beamline B18 at Diamond UK's National Synchrotron Light source and BM25 Spain Beamline at ESRF for their strong support in performing the experiments.
Authors would like to acknowledge the use of Servicio General de Apoyo a la Investigación-SAI, Universidad de Zaragoza.
Authors acknowledge the use of instrumentation as well as the technical advice provided by the National Facility ELECMI ICTS, node  Laboratorio de Microscopias Avanzadas (LMA)  at  Universidad de Zaragoza.

\section{Bibliography}
\bibliography{References}

\appendix
\section{Capacitive cantilever method}

For the cantilever structure and very thin film with a thickness below 100 nm, the central part of the reaction to the magnetoelastic stress is in the substrate since the elastic contribution of the films can be disregarded compared to that of the substrate. Therefore, 
the resulting variation of the cantilever radii of curvatures $R_{x}$ and $R_{y}$ is related to the magnetoelastic stress of the film. However, $\lambda$  is presented in some cases \cite{MARCUS199718}, but it requires knowledge of the \textit{c's}, which is difficult to obtain in some cases.

Considering the ME stress energy for cubic symmetry of the system studied here, which  is expressed in a Cartesian coordinate system as: 
\begin{equation}\label{emel}
e_{mel} = B_1\left[{
		\left({\alpha}_{z}^2-{\frac{1}{3}}\right)
		\left({\epsilon}_{zz}-{\frac{{\epsilon}_{xx}+{\epsilon}_{yy}}{2}}\right)
		+ {\frac{1}{2}}\left({\alpha}_{x}^2-{\alpha}_{y}^2\right)
		\left({\epsilon}_{xx}-{\epsilon}_{yy}\right)}\right]
	\nonumber 
\end{equation}
\begin{equation}	
	\\
	 +  2 B_2\left({ {\alpha}_{x} {\alpha}_{y} {\epsilon}_{xy} +{\alpha}_{y}
		{\alpha}_{z} {\epsilon}_{yz}+ {\alpha}_{z} {\alpha}_{x} {\epsilon}_{zx}}\right) 
\end{equation}
with
$\alpha 's$  cosines of the magnetization. In the case of the magnetization confined in the \textit{xy} plane: $\alpha_z =$ 0,  $\alpha_x = \cos\phi$, $\alpha_y
 = \sin\phi$, and
\begin{equation}
	e_{mel}(\phi)=  {\frac{1}{2}}B_1   
	\left({\epsilon}_{xx}-{\epsilon}_{yy}\right)\cos2\phi
	+ B_2 {\epsilon}_{xy} \sin2\phi + cte
\end{equation}

The expression for \textit{x} and \textit{y} parallel to FeGa[110] and [1-10] directions yields:
\begin{equation}
	e_{mel}(\phi)=  {\frac{1}{2}}B_2   
	\left({\epsilon}_{xx}-{\epsilon}_{yy}\right)\cos2\phi
	+ B_1 {\epsilon}_{xy} \sin2\phi + cte
\end{equation}

Therefore,  $e_{mel}$, with ${\epsilon}_{xx} \propto R_{x}^{-1} $, ${\epsilon}_{yy} \propto  R_{y}^{-1} $  and by energy minimization \cite{DUTREMOLETDELACHEISSERIE1994189}, the relationship between $B_1$ and $B_2$ with experiments performed in films with [100] and [110] axes along the cantilever edges are \cite{Koch1996, WEDLER2000896}: 
\begin{equation}
B_i= \frac{1}{6}E_{i} \frac{h^2_{subs}}{h_{film}}  \Delta \left(\frac{1}{R}\right) 
\end{equation}
with $\Delta(1/R)$ is the variation of radii of curvature with the magnetization oriented  
parallel ($\phi$ = 0) and transversely ($\phi$ = $\pi/2$) to the beam directions, with $h _{subs}$ and $h_{film}$ substrate and films thicknesses. The effect of the substrate elastic stiffness coefficients is included in the $E_i$ coefficient, a function of Young's modulus and Poisson ratio at each reference system. Several models \cite{MARCUS199718, Iannotti1999191} and finite elements calculations\cite{Watts97} have been discussed in the literature to determine the role of the transverse curvature of the beam. All of them provide values for $E_{i}$ as a function of the length/wide ratio but within the range determined by the free, $E_{free}$= E/(1+$\nu$), and flat, $E_{flat}$=E/(1-$\nu^2$), approximations for the transverse curvature. For the MgO(001),
and $B_1$, $E_{110}$ = 318.8 GPa  and  $\nu_{110} =$ 0.03 with $E \approx E_{flat} \approx E_{free}$. To determine $B_2$, $E_{100}$ = 250 GPa and $\nu_{100} =$ 0.24 with $E_{flat}$ = 264.5 GPa and $E_{free}$ = 201.8 GPa. For the aspect ratio of the cantilevers used here ($\approx$ 0.6), the value $E_2 =$ 215 GPa is obtained from ref. \cite{Watts97}.

The clamped  MgO substrate covers a fixed electrode entirely. 
Thus, the capacitance \textit{C} is given by $C = \epsilon_0 \int [D + \delta(x,y)]^{-1} dx dy  $
with the integral extended to the fixed capacitor's electrode area with length \textit{L} and width \textit{w}, D is the distance between parallel plates without magnetoelastic contribution,  $\delta(x,y)$ is the plate deflection at (x,y) coordinates and $\epsilon_0$ the vacuum permeability constant.
Because $\delta(x,y)$ is small the Taylor series approximation can be used.  

For a flat cantilever, $\delta(x,y) = 2x^2 R_x^{-1}$ with the capacitance variation $ \Delta C=-C^2_0(6\epsilon_0 f(A) R_x)^{-1}$ and f(A)= $ w (L_f-L_i)^2(L_f^3-L_i^3)^{-1}$ a function obtained considering the smaller capacitor electrode area. Boundaries for \textit{y} are $\pm$w/2 (w is the capacitor's smaller plate width) while for  \textit{x},  $L_i$ and $L_f=L_i+L$,  the distances between the clamping line and the capacitor fixed electrode edges.  All the films used in this study were grown on rectangular substrates 6 mm wide and 20 mm long. Thus, $L_f$ is about 9.8 mm, $L_i \approx$ 0.8 mm, and w = 4 mm for all samples. 

To estimate the role of the transverse curvature in the capacitance values,
 the transverse deflection of a free plate, $2y^2 R_y^{-1}$, is added to $\delta(x,y)$.   Its contribution to the flat cantilever capacitance is (considering $R_x=-R_y$) $ 0.25 w^2 (L_f-L_i)(L_f^3-L_i^3)^{-1}$, a correction as small as $\sim$ 0.038.  It decreases if $R_y$ evolves from infinite at the clamping line to the free end value at a certain large value of $L_f$.  
The small effect of the traverse curvature is clear because the sign of the capacitance variation jibes with the ME stress coefficient sign. Therefore, the variation of capacitance at the angular positions zero and $\pi/2$ in Figure 6 is linked to the ME coupling coefficient as:
\begin{equation}
B_i= E_{i} \frac{h^2_{subs}}{h_{film}}  \frac{\epsilon_0   f(A) }{C^2_0 }   \left[C(\pi/2)-C(0)\right]
\end{equation}

\end{document}